# Surface Effects on Anisotropic Photoluminescence in One-Dimensional Organic Metal Halide Hybrids

*Luke M. McClintock, Long Yuan, Ziyi Song, Michael T. Pettes, Dmitry Yarotski, Rijan Karkee, David A. Strubbe, Liang Z. Tan, Azza Ben-Akacha, Biwu Ma, Yunshu Shi, Valentin Taufour, and Dong Yu\**

L. M. McClintock, Z. Song, Y. Shi, V. Taufour, and D. Yu

Department of Physics and Astronomy, University of California-Davis, One Shields Avenue, Davis, CA 95616, USA

E-mail: yu@physics.ucdavis.edu

L. Yuan, M. T. Pettes, and D. Yarotski

Center for Integrated Nanotechnology, Los Alamos National Laboratory, Los Alamos, NM, 87545, USA

R. Karkee and D. A. Strubbe

Department of Physics, University of California, Merced, 5200 Lake Road, Merced, CA 95343, USA

L. Z. Tan

Molecular Foundry, Lawrence Berkeley Laboratory, 67 Cyclotron Rd, Berkeley, CA 94720, USA

A. Ben-Akacha and B. Ma

Department of Chemistry and Biochemistry, Florida State University, 600 W College Ave, Tallahassee, FL 32306, USA

**Abstract:** One-dimensional (1D) organic metal halide hybrids exhibit strongly anisotropic optical properties, highly efficient light emission, and large Stokes shift, holding promises for novel photodetection and lighting applications. However, the fundamental mechanisms governing their unique optical properties and in particular the impacts of surface effects are not understood. Here, we investigate 1D $C_4N_2H_{14}PbBr_4$ by polarization-dependent time-averaged and time-resolved photoluminescence (TRPL) spectroscopy, as a function of photoexcitation energy. Surprisingly, we find that the emission under photoexcitation polarized parallel to the 1D metal halide chains can be either stronger or weaker than that under perpendicular polarization, depending on the excitation energy. We attribute the excitation-energy-dependent anisotropic emission to fast surface recombination, supported by first-principles calculations of optical absorption in this material. The fast surface recombination is directly confirmed by TRPL measurements, when the excitation is polarized parallel to the chains. Our comprehensive studies provide a more complete picture for a deeper understanding of the optical anisotropy in 1D organic metal halide hybrids.

# 1. Introduction

Charge carriers and excitons in materials with weak electric field screening can interact strongly with the crystal lattice and be heavily dressed by phonons via a self-trapping mechanism, leading to lower mobility and large Stokes shifts.[1-3] Strong couplings between excitons and specific phonon modes have been directly observed, such as those seen in stack-engineered $WSe_2/MoSe_2$ heterojunction photodiodes,[4] monolayer $WSe_2$,[5] various forms of boron nitride,[6, 7] and chiral one-dimensional (1D) lead-free hybrid metal halides.[8] Halide perovskites and perovskite-related materials such as organic metal halide hybrids (OMHHs) have become one of the most promising classes of photoactive materials being studied, with potential applications in solar energy, photodetection, LEDs, and lasers. As a result of the formation of self-trapped excitons (STEs) with high binding energies, low-dimensional OMHHs exhibit bright white-light emission.[2, 9-11] A recent work demonstrated that the photoluminescence (PL) quantum yields can be further enhanced to up to 90% by high pressure.[12] This broadband and efficient emission, typically occurring only in materials with sufficient electron-phonon coupling, shows promise for uses in solid-state lighting devices.

Furthermore, 1D materials may exhibit strong optical anisotropy, offering additional exciting application opportunities. Fluorescence anisotropy has been observed in stretch-oriented poly(p-phenylenevinylene) (PPV),[13] J-aggregates of thiacarbocyanine dyes,[14] and 1D halogen-bridged Pt chains.[15] Anisotropic optical properties have been observed in single-walled carbon nanotubes through spectroscopic ellipsometry measurements.[16] Linear dichroism conversion has been reported in quasi-1D hexagonal perovskite chalcogenide $BaTiS_3$,[17] and the introduction of chiral cations in 1D perovskites has enabled direct detection of circularly polarized light.[18] A recent theoretical first-principles study confirmed the formation of STEs in 1D OMHHs $C_4N_2H_{14}PbX_4$ and established a polarization-luminescence relationship.[19] Experimentally, linear polarization-dependent PL has been reported recently in $C_4N_2H_{14}PbI_4$,[20] but how the PL linear polarization relates to the 1D metal halide chains of this 1D OMHH is not reported. More recently, a different 1D OMHH $C_3H_{10}NPbI_3$[21] has been reported to show PL linearly polarized along the 1D chain. However, this work applied only a fixed photoexcitation energy close to the exciton resonance energy of the 1D perovskite material and did not provide a clear physical mechanism accounting for the observed anisotropy. It is crucial to understand the fundamental mechanisms in anisotropic excitons and the self-trapping process in 1D OMHHs, to advance their potential broad applications in lighting, polarization optics, sensing, imaging, and communication technologies.

Here, we performed excitation energy- and polarization-dependent PL and time-resolved PL (TRPL) spectroscopy to better understand the optical anisotropy in the 1D single crystal OMHH, N,N'-dimethylethylenediammonium lead bromide ($C_4N_2H_{14}PbBr_4$ or DMEDAPbBr$_4$).[2] We first confirmed that the 1D metal halide chain is along the long axis of single crystal needle-like samples through rigorous X-ray diffraction (XRD) analysis. Then we performed comprehensive PL spectroscopy with the photoexcitation wavelength ranging from 360 nm to 410 nm at both room temperature and 8 K. Two linear polarizers were used to independently control and analyze both excitation and emission polarization. Surprisingly, we found that the PL reaches a maximum (minimum) when photoexcitation is linearly polarized perpendicular to the 1D metal halide chain when the excitation wavelength is shorter (longer) than 380 nm, respectively. The ratio of PL emission under different excitation polarizations ($I_{max}/I_{min}$, where $I_{max}$ and $I_{min}$ are the maximum and minimum PL emission intensities, respectively) also sensitively depends on the excitation energy and reaches up to 9.4. TRPL measurements showed an additional linearly polarized fast decay component (< 20 ps) when photoexcitation was parallel to the 1D chain. We carried out first principles calculations that showed highly anisotropic optical absorption at low excitation energy. Based on these experimental and theoretical results, we attribute the strong excitation energy dependent anisotropic emission to fast recombination at the surface of 1D OMHHs. Our work highlights the importance of surface effects on optical properties of the 1D OMHHs.

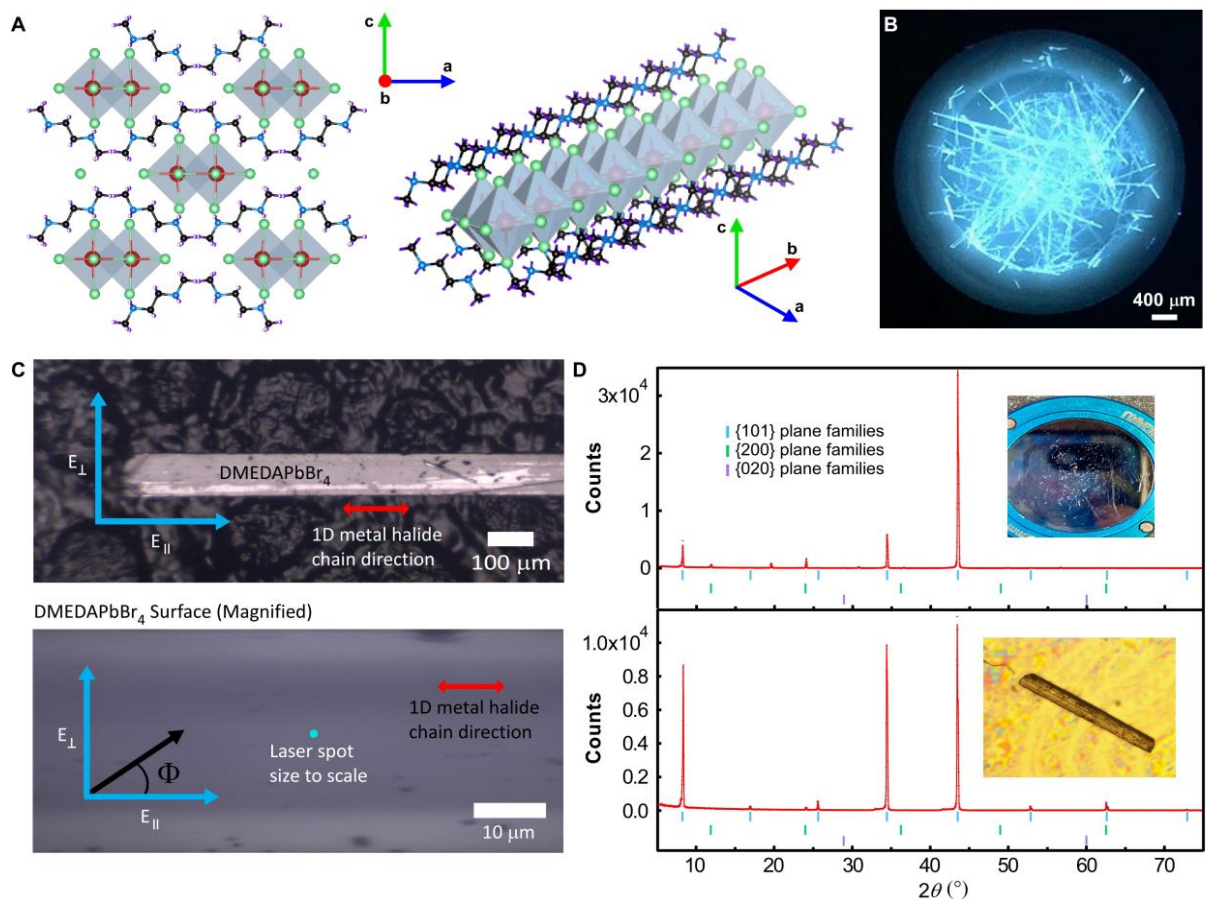

**Figure 1.** A) Structure of DMEDAPbBr$_4$ that can be described as lead bromide quantum wires wrapped by the organic cations (red spheres: lead atoms; green spheres: bromine atoms; blue spheres: nitrogen atoms; black spheres: carbon atoms; purple spheres: hydrogen atoms; grey polyhedra: PbBr$_6^{4-}$ octahedrons). Hydrogen atoms are hidden for clarity in the image on the right. The red $\vec{b}$ axis indicates the 1D chain direction. B) Single crystals of DMEDAPbBr$_4$ under UV light (365 nm). C) A single DMEDAPbBr$_4$ sample on carbon tape with 1D metal halide chain and laser polarization directions labeled. D) XRD results for many crystals (top) as well as an isolated crystal (bottom).

## 2. Results

### 2.1. Synthesis and Crystal Structure

The synthetic method of DMEDAPbBr$_4$ can be found in the Methods section. The crystal structure of DMEDAPbBr$_4$ can be understood by 1D chains surrounded by organic cations to form core-shell-like quantum wires (**Figure 1A**). The needle-like single crystals are about a few mm long and 100 μm wide and tall as shown in Figure 1B and C. To determine the orientation of the 1D metal halide chains, we performed careful XRD analysis on single crystal needles lying on the substrates. The orthorhombic crystal has similar lattice constants along the a and c axes (a = 14.62 Å and c = 14.41 Å), but the lattice constant along the b axis (the 1D metal halide chain) is much shorter (b = 6.10 Å). The crystal orientation can be identified from the missing XRD peaks since the diffraction only occurs in the crystal planes parallel to the substrates. As shown in Figure 1D, top, the diffraction pattern from an ensemble of needles exhibits the {101} and {200} plane families, in which the b index is always zero. The {020} plane family is labeled (purple markers) in Figure 1D to demonstrate the absence of {020} peaks. The intensities of the {101} peaks are larger, which indicates that the needle cross sections are likely terminated by the {101} planes. This is confirmed by performing the measurement again on a single needle (Figure 1D, bottom) which shows only the {101} peaks. This clearly indicates that the b axis must be along the long axis of the needle, which is always parallel to the substrate. Otherwise, diffraction peaks with nonzero b index are expected to show up because of the random orientation of the short axis of the needles. We note that there is an additional small peak around 24° that could correspond to the {400} peak. This is likely due to a tiny fragment of crystal that has chipped off and is aligned in a different direction.

## 2.2. Photoluminescence

The experimental setup is shown in **Figure 2A**, where a linearly polarized tunable UV laser is focused onto the 1D perovskite crystals with normal incidence and a spot size of about 1 µm. Thanks to the tight laser focus and high optical image resolution, we are able to measure completely clean regions of the crystal surface (Figure 1C, bottom). The PL results at various spots of the samples are consistent and do not vary much as long as the photoexcitation location is free of debris and obvious defects. The linear polarization of the photoexcitation can be adjusted by a half waveplate (designed specifically for the UV range). Another linear polarizer for the visible light can also be inserted in front of the streak camera to study the polarization of the emitted light. We have observed that the overall PL intensity (with the linear polarizer in front of the streak camera removed) sensitively depends on both the polarization of the excitation laser beam as well as the wavelength (Figure 2B). Under 360 nm excitation, when the electric field of the incident laser beam is perpendicular to the 1D metal halide chains the emission intensity reaches a maximum, which, for the sample shown in Figure 2B, is about 5 times stronger than the intensity when the field is parallel. Surprisingly, this trend is reversed at lower photoexcitation energy. Under 400 nm excitation, the PL emission maximum is reached when the laser polarization is parallel to the 1D chain. This anisotropic emission intensity can be well-fit with the equation $I(\Phi) = I_0 \cos[2(\Phi - \Phi_0)] + K$, where $\Phi$ is the angle between the photoexcitation polarization and the 1D chain, $\Phi_0$ is the phase shift, $I_0$ is the polarization-dependent intensity amplitude, and $K$ is the polarization-independent intensity component[20] (Figure 2B). $\Phi_0$ is found to be close to 0 for 400 nm and π/2 for 360 nm from the fitting. The reversal of polarization dependence of PL upon increasing the photoexcitation energy has not been reported to our knowledge. The 400 nm excitation is close to the minimum absorption energy of $DMEDAPbBr_4$, which corresponds to the exciton resonance energy. We plot the total PL counts under photoexcitation with polarization parallel and perpendicular to the chain at various excitation wavelengths in Figure 2C and Figure S2 (for a different sample) in the Supporting Information. The PL intensity first increases and then decreases as the excitation wavelength increases from 360 nm to 400 nm. There is also a clear cross-over at 380-390 nm, where the perpendicular polarization generates stronger PL at shorter wavelength.

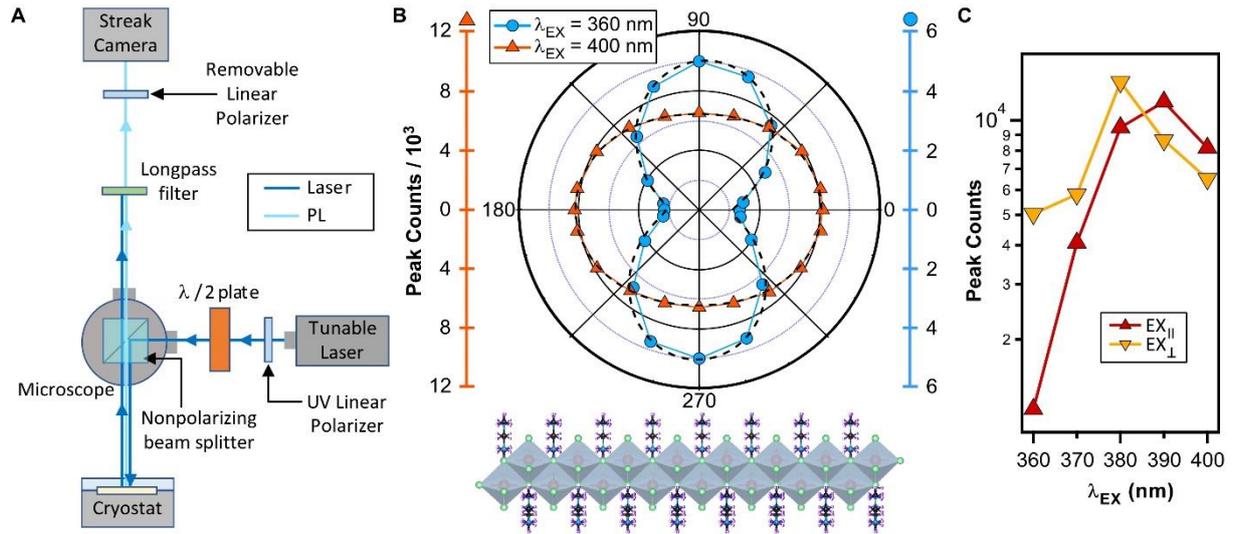

**Figure 2.** A) Schematic of experimental setup for polarization-dependent PL and TRPL measurements. B) Polar plot of peak PL counts as a function of excitation polarization angle $\Phi$ for high (360 nm) and low (400 nm) energy excitation. The dashed lines are curve fits. Polarization at 0° is parallel to the 1D chain of $DMEDAPbBr_4$. C) Total PL emission counts as a function of excitation wavelength and orthogonal excitation polarizations. All PL counts are normalized to photoexcitation power. Data is collected from Sample #3.

The PL emission spectra are broad, but their shape is independent of the excitation polarization (**Figure 3A**) and excitation wavelength for both 295 K and 8 K. The broadband emission has been attributed to the exciton self-trapping.[2] The total emission is stronger and the spectral width is narrower at low temperature. We define an excitation anisotropy ratio as $R_{EX} = I_{EX\perp}/I_{EX\parallel}$, where $I_{EX\parallel}$ ($I_{EX\perp}$) is the peak PL count (with the detector polarizer removed) when photoexcitation is linearly polarized parallel (perpendicular) to the 1D chain, respectively. $R_{EX}$ increases at lower temperature from 4.3 to 9.5 at $\lambda_{EX}$ = 360 nm for Sample #3. Curiously, at low temperature, $R_{EX}$ abruptly increased between 370 nm and 390 nm excitation. The reason behind this bump in the trend is currently unknown but may be caused by fine electronic structure revealed at low temperature. $R_{EX}$ varies from sample to sample, up to 9.4 at room temperature for one sample (Figure 3B). The variation is presumably due to the alignment of 1D chains that can sensitively depend on the growth conditions. As excitation wavelength increases, $R_{EX}$ decreases and then crosses below 1 around 380 nm (Figure 3B), meaning parallel excitation now causes brighter emission. This trend is repeatable across multiple DMEDAPbBr$_4$ samples, with the cross-over excitation wavelength varying between about 375 and 385 nm. The anisotropy ratio is also fairly large at long wavelength, up to 1/3.

The PL emission is highly linearly polarized parallel to the 1D chain, regardless of the excitation polarization, as shown in Figure 3C. We define an emission anisotropy ratio as $R_{EM} = I_{EM\parallel}/I_{EM\perp}$, where $I_{EM\parallel}$ ($I_{EM\perp}$) is the peak PL count when the detector polarizer is parallel (perpendicular) to the 1D chain, respectively. $R_{EM}$ are extracted from Figure 3C to be 11.4 (11.2) for perpendicular (parallel) excitation at room temperature, corresponding to a degree of polarization ($P = [I_{EM\parallel} - I_{EM\perp}] / [I_{EM\parallel} + I_{EM\perp}]$) of 84%. These values are significantly higher than those reported in 1D polymer and Pt chains,[13, 15] as well as the previously reported values in 1D OMHHs C$_4$N$_2$H$_{14}$PbI$_4$ and C$_3$H$_{10}$NPbI$_3$,[20, 21] demonstrating the strong anisotropy and high quality of our samples. The PL peak position shifts from 510 nm for the emission parallel to the chain to 485 nm for that perpendicular to the chain (Figure 3C). A similar blue-shift of PL spectra for transverse polarization has been observed in C$_3$H$_{10}$NPbI$_3$.[21] This spectral shift indicates the transverse exciton state has a slightly higher energy than the longitudinal exciton state.[22, 23]

## 2.3. Time-Resolved Photoluminescence

PL emission under a pump pulse (< 200 fs) at 380 nm shows a very slow decay (10s of ns) under excitation perpendicular to the 1D chain, consistent with previous reports.[2] However, an initial sharp decay (< 20 ps) has been observed when the excitation polarization is parallel to the 1D chain (**Figure 4A**). The presence and magnitude of this sharp decay feature appear to depend on both excitation polarization as well as excitation wavelength. At 360 nm excitation the sharp feature is relatively weaker than at 380 nm excitation under parallel excitation (Figure S3). Under all excitation wavelengths, DMEDAPbBr$_4$ appears to emit both parallel and perpendicular to the 1D chains, but the parallel component is always significantly stronger. The initial sharp decay peak is only observed when the detector polarizer is parallel to the chain in the sample shown in Figure 4A, though a small peak is also observed for perpendicular emission polarization in another sample (Figure S4). The PL emission spectral peak blue-shifts from 530 nm to 500 nm as the PL intensity increases to maximum in about 40 ps after the pump pulse (Figure 4B). Then the PL falls to a slow decay state in about 50 ps. The measured fast decay time stays at about 12-14 ps (Figure S5 and S6) when the pump intensity increases by 20 times at both 295 K and 11 K, though the measured decay time is likely limited by the temporal resolution of the setup (Figure S1) and should be considered as the upper bound of the real value. The PL counts for both the short and long components are linear with the fluence of the pump laser (Figure S5 and S6), at both 295 K and 11 K. The slow decay component is at least tens of nanoseconds, which cannot be accurately determined in our experiment because of the high repetition rate of the excitation laser pulses.

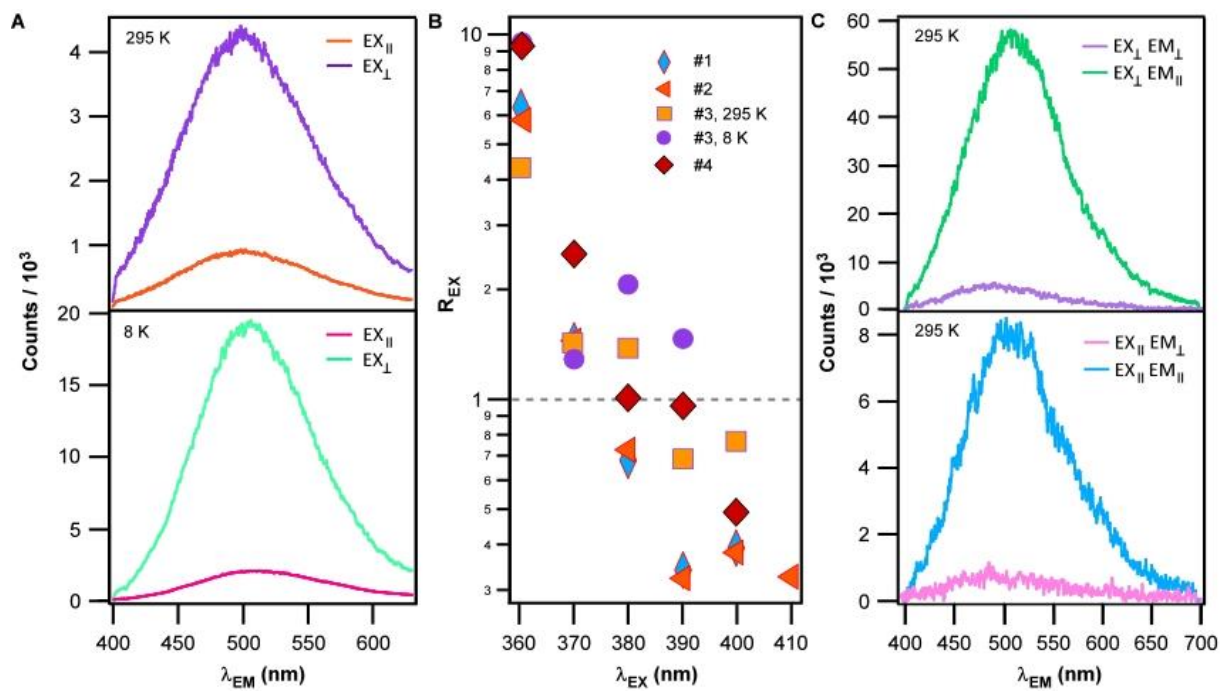

**Figure 3.** A) Excitation polarization-dependent PL spectra for Sample #3 at room temperature and 8 K, respectively. Counts drop off at 400 nm due to a long pass filter. The measurements were performed with excitation wavelength of 360 nm and no polarizer was in the detection path. B) $R_{EX} = I_{EX\perp}/I_{EX\parallel}$ for representative DMEDAPbBr$_4$ samples at room temperature and 8 K. Fewer data points are displayed at 410 nm excitation due to the PL becoming too weak for accurate measurements. C) Room temperature PL emission spectra for Sample #2 at 360 nm excitation with the detector linear polarizer inserted.

## 3. Discussion

We first summarize the main observations: (a) PL emission is stronger when excitation polarization is parallel (perpendicular) to the 1D chains at $\lambda_{EX}$ = 400 nm (360 nm), respectively. (b) PL emission is always more strongly polarized along the chain, regardless of excitation wavelength or polarization. (c) TRPL shows a fast decay (< 20 ps) only when excitation polarization is parallel to the chain. All these observations can be understood by the surface effects on PL as detailed below.

The optical absorption of 1D halide perovskites is expected to be highly anisotropic, strongly depending on the optical polarization. This is because the optical electric field parallel to the chain can more easily polarize the electrons along the 1D chain composed of inorganic octahedra. Confined by the insulating organic structure surrounding the chain, electrons are harder to be polarized by transverse electric field. To confirm, we performed plane-wave DFT calculations in the Quantum ESPRESSO code,[24] with the Perdew-Burke-Ernzerhof (PBE)[25] exchange-correlation functional. Relaxation starting from the XRD structure ($a$ = 14.62 Å, $b$ = 6.10 Å, $c$ = 14.40 Å, α = β = γ = 90°) gave lattice parameters in close agreement ($a$ = 14.70 Å, $b$ = 6.06 Å, $c$ = 14.56 Å, α = β = γ = 90°), as in work on other hybrid perovskites and OMHHs.[26, 27] Our calculations use a primitive cell of this body-centered tetragonal structure, with lattice parameters $a$ = $b$ = $c$ = 10.80 Å, and α = 32.48°, β = 94.86°, γ = 85.88° with the Pb-Br chain along the $z$ direction (see Supporting Information). The calculated electronic band structure of the 1D metal halides in the GW approximation (with a simple spin-orbit correction) is shown in **Figure 5A**; the gap is indirect and computed to be 3.6 eV, though the difference between the direct and indirect bandgap is only 16 meV. The valence band maximum (VBM) has contributions mostly from the p-orbitals of Br, whereas at the conduction band minimum (CBM), p-orbitals of Pb dominate (Figure 5C). The electronic bands are dispersive along the Pb-Br chain direction but are nearly flat along the other two perpendicular directions.

Additionally, we calculated the absorption spectrum through the Bethe-Salpeter Equation (BSE) in the BerkeleyGW code,[28] shown in Figure 5B. Results from the random phase approximation (RPA) are also provided for comparison, which do not include excitonic effects. By comparing the absorption peaks of BSE and RPA, we find a substantial exciton binding energy (830 meV), as is usual in low-dimensional structures.[29] The first absorption peak is strong along the Pb-Br chain (z-polarized), consistent with the Pb-Br chain mainly contributing to the electronic transition at the band edges. The peak is mainly due to transitions at $\vec{k}$ = Z from the VBM and the next band below the VBM to the CBM and the next band above the CBM (CBM+1). For x- and y-polarizations, the dominant transitions are at $\vec{k}$ = X and Y respectively, and in both cases are from VBM to CBM and CBM+1. Figure 5B shows the large anisotropy in absorption in this material.

The above calculation clearly indicates strong anisotropic absorption for the lowest energy excitation, and hence in the excitation energy range used in our experiment (3.02-3.44 eV), which is just above the (perhaps slightly underestimated) calculated peak energy. The calculated optical transition rates are much higher for the parallel electric field in the lowest energy peak shown in Figure 5B. The corresponding absorption depth reaches as shallow as 10 nm for polarization parallel to the Pb-Br chain, at least two orders of magnitude smaller than that under perpendicular polarization (Figure 5D). The sharp rise of the peak on the low energy side agrees with the rapid increase in the optical absorption as the excitation energy increases within the experimentally accessible range. Consequently, at $\lambda_{EX}$ = 400 nm, the photon energy is just above the minimum absorption energy and the absorption is weak. The photoexcited charge carriers are located relatively deep in the bulk (depicted in Figure 4C). In this case, the absorption is enhanced when the polarization is parallel to the chain as the perpendicularly polarized light is likely only absorbed partially by the approximately 100 μm thick samples. This leads to a moderately enhanced emission under parallel polarized excitation. In contrast, at $\lambda_{EX}$ = 360 nm, the absorption is much stronger due to resonance with the lowest excitonic level and the photoexcited carriers are generated closer to the surface. For parallel polarization, the absorption is even stronger, leading to a very shallow distribution of photoexcited carriers. Under these conditions, the PL intensity is strongly reduced because of fast recombination at the surface. At $\lambda_{EX}$ = 380 nm, both perpendicular and parallel excitations generate about the same PL intensities, when the surface effects under parallel polarization and the incomplete absorption under perpendicular polarization have similar impacts to the PL reduction.

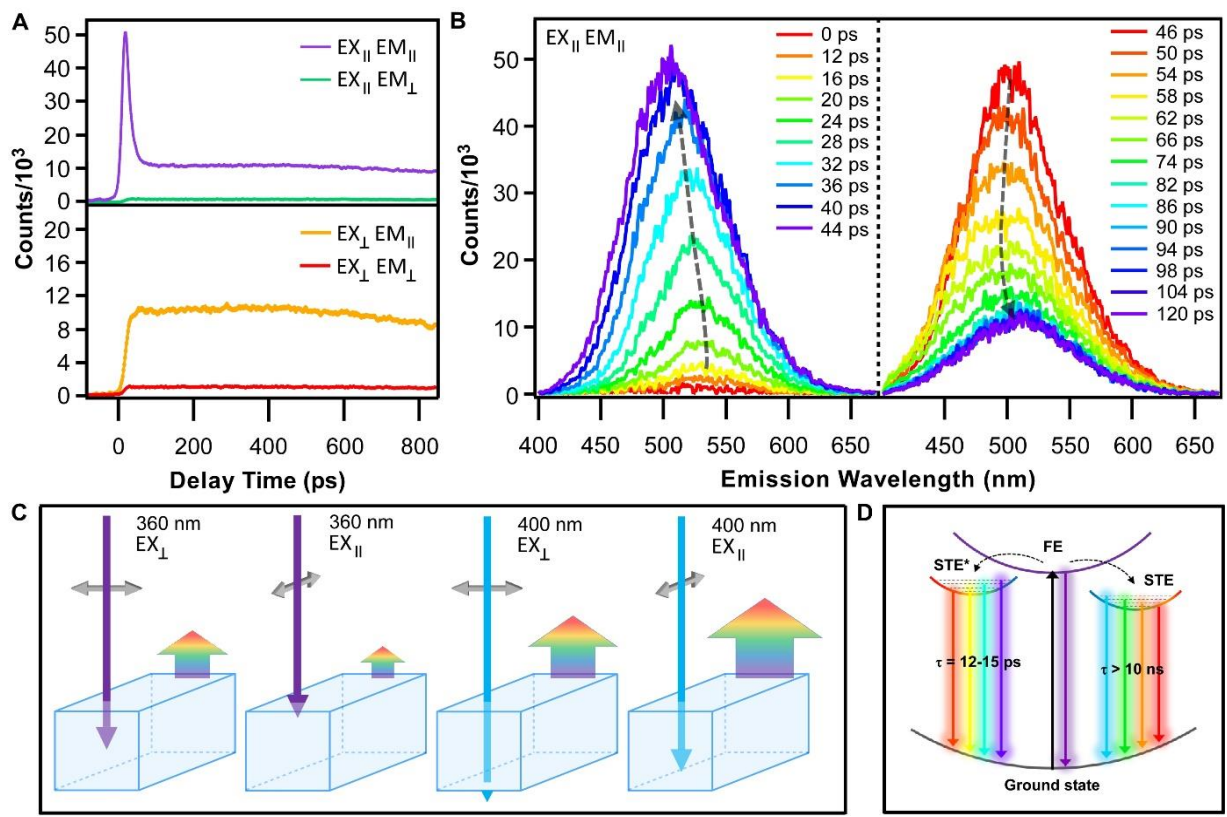

**Figure 4.** A) PL counts as a function of delay time after 380 nm pulsed excitation at room temperature (Sample #5) with different excitation and emission polarization configurations. B) PL emission spectra at various delay times across the rise and fall of the sharp feature seen under parallel excitation and emission in A). C) Proposed surface effect model. At 360 nm, parallel polarization is absorbed very close to the sample surface, resulting in reduced PL emission. At 400 nm, both polarizations can now penetrate farther into the bulk, but the parallel polarization is more readily absorbed, resulting in relatively stronger emission. D) Energy diagram and relaxation pathways for the proposed TRPL surface effect mechanism. STEs* are created near the surface and recombine fast, while STEs are in the bulk and recombine slowly.

Regardless of excitation polarization or energy, the emission is mostly polarized parallel to the 1D chain, indicating the transition dipoles of self-trapped excitons are preferentially parallel to the 1D chain. However, there is always a small perpendicular component of emission as well. This may simply be caused by the non-perfect alignment of the 1D chains, though the sharp XRD peaks indicate the samples are highly crystalline. Another possibility is that a small portion of self-trapped excitons exist with transition dipoles not entirely parallel to the chain. For example, the transition dipole moment has been estimated to make an angle of 65-70° with the chain direction in J-aggregates of thiacarbocyanine dyes.[14]

TRPL results can also be understood with the above hypothesis of surface effects. We first highlight the key observations in TRPL measurements in detail below: (i) parallel excitation results in two distinct decay constants with a fast component shorter than 20 ps and a slow component longer than 10 ns, while perpendicular excitation only produces the slow component; (ii) the initial sharp PL decay is highly linearly polarized (Figure 4A); (iii) both the fast and slow components are linear with the fluence of the pump laser (Figure S5 and S6) at both 11 K and 295 K; (iv) the emission spectra of the fast and slow components have a similar shape, with the fast component slightly blue-shifted. Observation (i) indicates that the initial fast recombination is likely induced by the excitons generated close to the surface under parallel excitation. Observation (ii) implies that the emission from trapped charge at the radiative surface defect states is unlikely to account for the initial fast PL decay, since the emission from surface defects is expected to have random polarization. Observation (iii) indicates that the initial sharp decay is unlikely to be caused by a nonlinear effect such as Auger recombination. Lastly, the similarity in the broad emission spectra in observation (iv) suggests that both the fast and slow components are created by a similar recombination process of STEs. The different decay times and emission energies are likely caused by the types of STEs generated under different polarizations as depicted in Figure 4D. Under parallel excitation, a significant portion of STEs is created near the surface (labeled as STE* in Figure 4D). These STEs* recombine at a faster rate with slightly higher emission energy.

We briefly discuss the possible mechanisms that may account for the different dynamics of STEs* near the surface. The slow recombination in 1D metal halides has been attributed to spin-forbidden transitions of STEs.[1, 30] The fast PL component is most likely caused by nonradiative recombination via surface defects, such as dangling bonds or vacancies. This provides a fast nonradiative pathway, substantially reducing the overall PL intensity under parallel polarization. The observed initial fast PL component is still from the radiative STE* recombination, though only a small portion of STE*s recombine radiatively, as most excitons near the surface recombine nonradiatively through the surface defects. The excitons generated in the bulk away from the surface do not suffer from this fast defect-mediated nonradiative recombination, as the interchain diffusion of excitons is expected to be very slow. However, we do not exclude other possible mechanisms expediting the STE* recombination near the surface. The proximity to the surface may help mix the single and triplet states to allow faster recombination. Singlet-triplet conversion through intersystem crossing (ISC) has been reported recently[31-33] and ISC can be induced through crystal structure distortion by stress.[34] We speculate that surface reconstruction may have a similar effect. The out-of-plane electric field at the sample surface may also have an impact on the charge recombination dynamics. The ion migration in 1D OMHHs[35] may create surface charge that results in an electric field near the surface. By comparing PL by one-photon and two-photon excitation, it has been observed that the recombination lifetime is significantly reduced and the emission spectra blue-shift near the surface of MAPbBr$_3$,[36] similar to our observations in Figure 4A and B. Further work such as magnetic field- and electric field-dependent TRPL measurements, as well as passivation of surface defects by coating, are needed to fully clarify the exact mechanism.

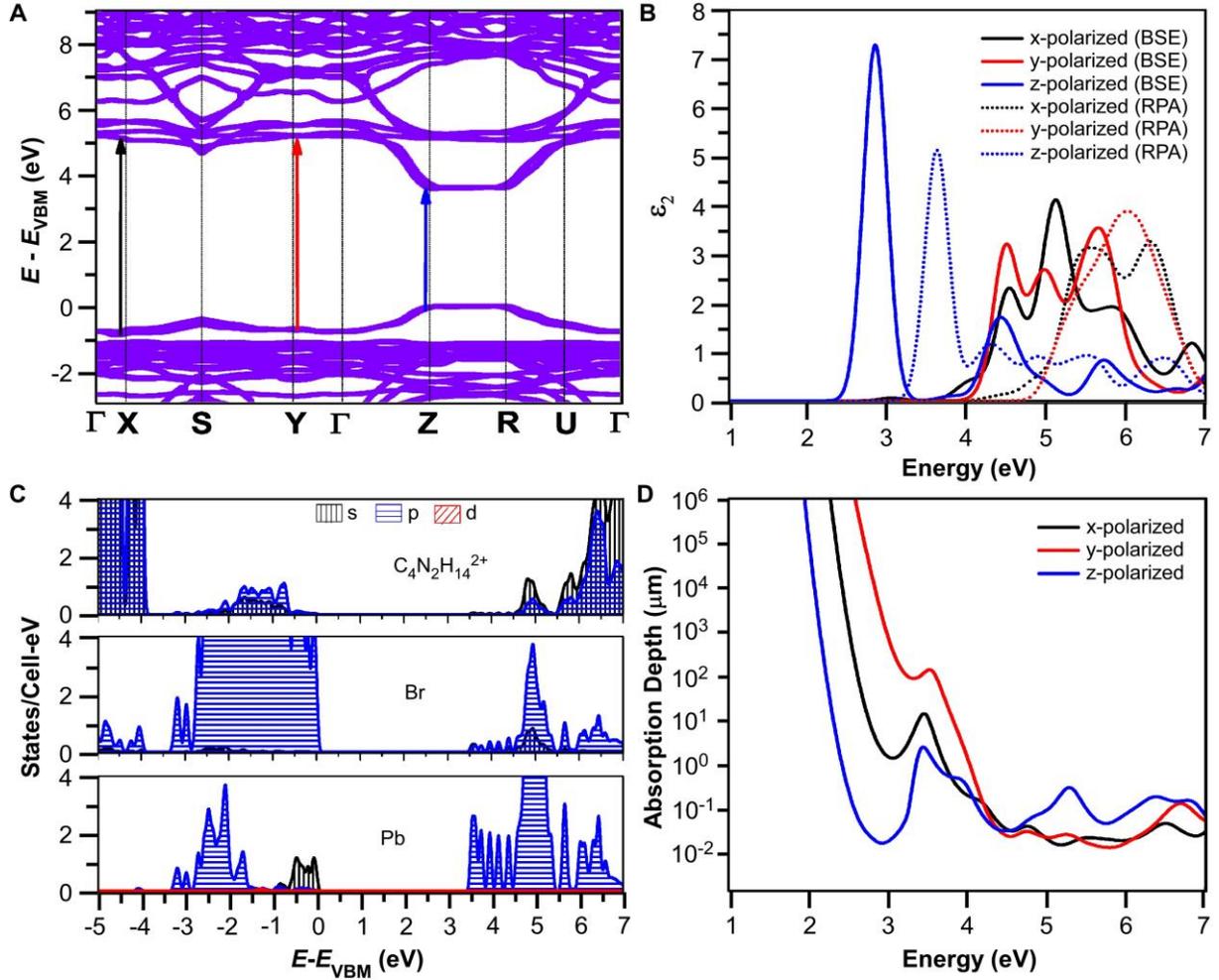

**Figure 5.** Electronic structure simulations. A) Electronic band structure, in the conventional Brillouin zone, in the GW approximation plus a spin-orbit correction to the gap. Arrows show the dominant transitions for the lowest-energy peaks for light polarized along x (black), y (red), and z (blue) directions, where z is the direction along the Pb-Br chains. B) Polarized absorption spectra with (BSE, solid) and without (RPA, dashed) electron-hole interactions, based on the GW bandstructure with spin-orbit correction to the gap. C) Partial density of states from DFT to show atomic orbital character of bands, with gap corrected by GW and spin-orbit coupling. D) Absorption depth for polarized light, from BSE as in B).

## 4. Conclusion

In summary, we performed excitation energy- and polarization-dependent PL and TRPL spectroscopy to elucidate optical anisotropy properties in the 1D single crystal OMHH, N,N'-dimethylethylenediammonium lead bromide. We confirmed the parallel orientation of the 1D metal halide chain with respect to the long axis of the needle-like crystals. We performed comprehensive polarization-dependent PL and TRPL spectroscopy with the photoexcitation wavelength ranging from 360 nm to 410 nm at both room temperature and 8 K. Interestingly, we found that the trend of anisotropic PL reversed as we increased photoexcitation energy, where perpendicular (parallel) excitation resulted in more emission at higher (lower) energy, respectively. The anisotropy ratio, defined by $R_{EX} = I_{EX\perp}/I_{EX\parallel}$, increased from 1/3 at 410 nm to 9.4 at 360 nm. Curiously, TRPL measurements revealed an additional fast decay component of < 20 ps only when excitation was parallel to the 1D chains, which could be caused by the surface impacts on the recombination of self-trapped excitons. We emphasize that the fast PL component is linearly polarized and exhibits an emission spectrum similar to the slow component, so it is unlikely to be induced by radiative recombination via surface defects. Instead, we speculate that the fast recombination is caused by nonradiative recombination via surface defects or by the crystal structure distortion near the surface, which mixes the triplet and singlet states of STEs. Finally, we carried out first principles calculations that displayed highly anisotropic optical absorption in the range of our excitation energies. Based on these experimental and theoretical results, we attribute the strong excitation energy dependent anisotropic emission to a drastically different recombination process at the surface of 1D OMHHs. This work highlights the importance of surface effects in understanding the optical properties of 1D OMHHs and utilizing their anisotropic and broadband emission in novel applications for lighting and photodetection.

## 5. Methods

### 5.1. Sample Synthesis and Characterization

Lead(II) bromide (PbBr$_2$, 99.999%), N,N'-dimethylethylenediamine (DMEDA, 98%), hydrobromic acid (HBr, 48 wt.% in H$_2$O) were purchased from Sigma-Aldrich. Acetone (≥ 99.5% ACS) was purchased from VWR Chemicals BDH. All reagents and solvents were used without further purification unless otherwise stated. Equimolar amounts of lead(II) bromide and N,N'-dimethylethylenediamine (0.27 mmol) were added into 10 mL of 48 wt.% hydrobromic acid, then the mixture was sonicated to yield a clear solution. Needle-like single crystals of C$_4$N$_2$H$_{14}$PbBr$_4$ were obtained through diffusion of acetone into a 1 mL of the precursor solution for 24 h. The crystals were washed with acetone and then dried under reduced pressure for further use and characterization.

The orientation of the single crystals was investigated with X-ray diffraction (Rigaku MiniFlex600 Diffractometer).[37] A C$_4$N$_2$H$_{14}$PbBr$_4$ single crystal was placed on the substrate with one facet facing upward for the X-ray 2θ-scan, as is shown in Figure 1D. The facet of the single crystal was identified with the group of diffraction peaks. An ensemble of crystals was also checked by placing multiple crystals on the puck with one of their side facets parallel to the substrate. The result shows that only the {101} and {200} peaks were observed, indicating that the side facets of the crystals belong to either the {101} or the {200} plane families.

### 5.2. Photoluminescence Measurements

An ultrafast Ti:Sapphire pulsed laser (Coherent Chameleon Ultra) with a tunable wavelength range of 680 - 1080 nm and pulse width of 140 fs was used for excitation. For the high energy excitation, the Chameleon output was sent through a frequency doubling module. Collected photons are sent to a spectrometer (Acton, Princeton Instruments) and a Hamamatsu streak camera. The average laser intensity typically varies from about 10 to 50 W/cm$^2$ across the low temperature and room temperature measurements. Excitation polarization is controlled via a combination of a UV linear polarizer, UV half-wave plate, and nonpolarizing cube beam splitter. Emission polarization is measured via another linear polarizer in line with the detection path. The low temperature PL measurements were carried out in a close-cycle cryostat (Montana Instruments).

### 5.3. Simulation and Modeling

ONCV pseudopotentials[38] from PseudoDojo[39] were used. Most calculations were scalar relativistic. A wavefunction energy cutoff of 816 eV and a 3 × 3 × 3 half-shifted $k$-grid were used for self-consistent field (SCF) calculations. Forces and stresses were relaxed below $10^{-4}$ Ry/bohr and 0.1 kbar, respectively. Density of states calculations used a 20 × 20 × 20 half-shifted *k*-grid and a broadening of 0.05 eV. For BerkeleyGW calculations, we found that 300 empty bands, a 4 × 4 × 4 *q*-grid, and a 204 eV screened-Coulomb cutoff converged GW quasiparticle corrections near the gap to within 100 meV. Optical absorption spectra computed with the Bethe-Salpeter equation in BerkeleyGW use 18 occupied states and 24 unoccupied states and a 8 × 8 × 8 fine *k*-grid, and are plotted with 0.1 eV Gaussian broadening. To obtain a simple spin-orbit correction, fully relativistic DFT calculations were performed on the same structure. The bandstructure had a gap reduced by 0.5 eV. This difference was then applied as a rigid shift to the GW, RPA, and BSE results, motivated by the typical finding that energy shifts due to spin-orbit are similar in DFT and GW.[40] Similarly, a rigid shift to the DFT PDOS was applied with the sum of the gap corrections from GW (+1.30 eV) and spin-orbit coupling.

**Supporting Information**

Supporting Information is available from the Wiley Online Library or from the author.
Supporting Information Available:
Temporal resolution calibration of the TRPL setup; energy- and polarization-dependent PL results from another sample; wavelength-, power-, temperature-dependence of TRPL results; coordinate file for DFT-relaxed structure of primitive unit cell.

**Acknowledgements**

We thank Henry Travaglini, Tugrul Senger, Prashant Padmanabhan, and Jake Pettine for assistance in additional sample measurements and discussion. This work was supported by the U.S. National Science Foundation Grants DMR-2209884 and DMR-2105161. L. M. acknowledges the DOE SCGSR fellowship and UC-National lab in-residence graduate fellowship. Y. S and V. T. acknowledge support from the UC Lab Fees Research Program (LFR-20-653926). A. B-A. and B. M. thank support from NSF grant DMR-2204466. L. Z. T was supported by the Molecular Foundry, a DOE Office of Science User Facility supported by the Office of Science of the U.S. Department of Energy under Contract No. DE-AC02-05CH11231. The TRPL work was performed at the Center for Integrated Nanotechnologies, an Office of Science User Facility operated for the U.S. Department of Energy (DOE) Office of Science. Los Alamos National Laboratory, an affirmative action equal opportunity employer, is managed by Triad National Security, LLC for the U.S. Department of Energy's NNSA, under contract 89233218CNA000001. Computational work by R. K. and D. A. S. was supported by the Air Force Office of Scientific Research under award number FA9550-19-1-0236 and used resources of the Multi-Environment Computer for Exploration and Discovery (MERCED) cluster at UC Merced, funded by National Science Foundation Grant No. ACI-1429783 and the National Energy Research Scientific Computing Center (NERSC), a U.S. Department of Energy Office of Science User Facility operated under Contract No. DE-AC02-05CH11231.

## Conflict of Interest

The authors declare no conflict of interest.

## Data Availability Statement

The data that support the findings of this study are available on request from the corresponding author.

## ORCID

Dong Yu: 0000-0002-8386-065X

Valentin Taufour: 0000-0002-0024-9960

Luke M. McClintock: 0000-0003-1206-4645

David A. Strubbe: 0000-0003-2426-5532

Rijan Karkee: 0000-0003-0124-5213

Liang Z. Tan: 0000-0003-4724-6369

Michael T. Pettes: 0000-0001-6862-6841

**Table of contents entry**: One-dimensional (1D) organic metal halide hybrids exhibit highly anisotropic optical properties and have promising potentials in novel photodetection and lighting applications because of their quantum confined 1D molecular chains. We show that surface effects play an important role in the light emission process in these materials through energy- and polarization-dependent photoluminescence measurements.

*Luke M. McClintock, Long Yuan, Ziyi Song, Michael T. Pettes, Dmitry Yarotski, Rijan Karkee, David A. Strubbe, Liang Z. Tan, Azza Ben-Akacha, Biwu Ma, Yunshu Shi, Valentin Taufour, and Dong Yu\**

**Surface Effects on Anisotropic Photoluminescence in One-Dimensional Organic Metal Halide Hybrids**

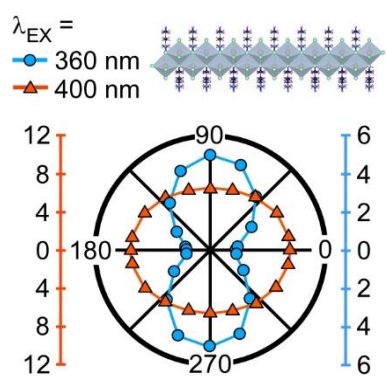

**Surface Effects on Anisotropic Photoluminescence in One-Dimensional Organic Metal Halide Hybrids**

*Luke M. McClintock, Long Yuan, Ziyi Song, Michael T. Pettes, Dmitry Yarotski, Rijan Karkee, David A. Strubbe, Liang Z. Tan, Azza Ben-Akacha, Biwu Ma, Yunshu Shi, Valentin Taufour, and Dong Yu\**

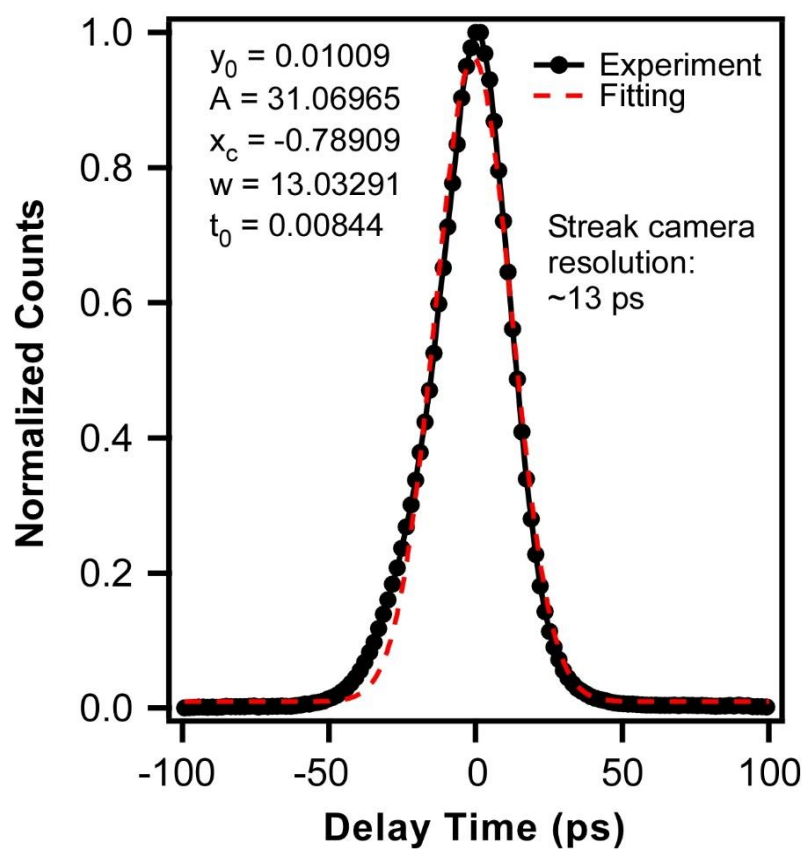

**Figure S1.** Streak camera temporal resolution calibration, obtained by measuring the input laser pulse. Fitting equation: $y = y_0 + (A/t_0) \exp[0.5(w/t_0)^2 - (x-x_c)/t_0][\text{erf}(z/\sqrt{(2)})+1]/2$, where $z = (x - x_c)/w - w/t_0$. The fitting parameter $w$ is the estimated temporal resolution in ps.

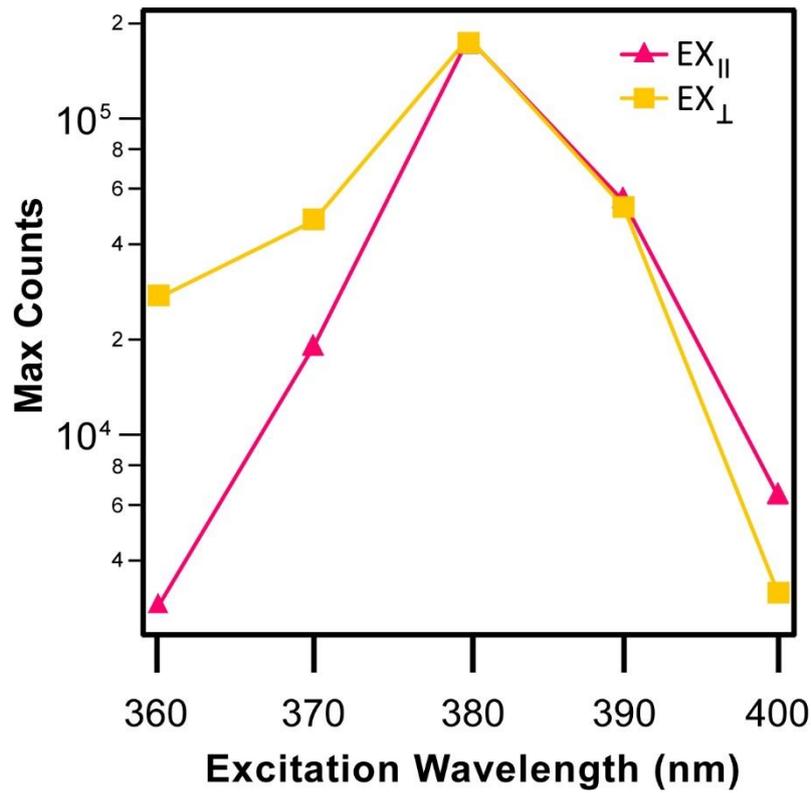

**Figure S2.** Total PL emission counts as a function of excitation wavelength and orthogonal excitation polarizations. All PL counts are normalized to photoexcitation power, measured in Sample #4. Note that the trend is similar to the data collected from Sample #3, as shown Figure 2C in the main text.

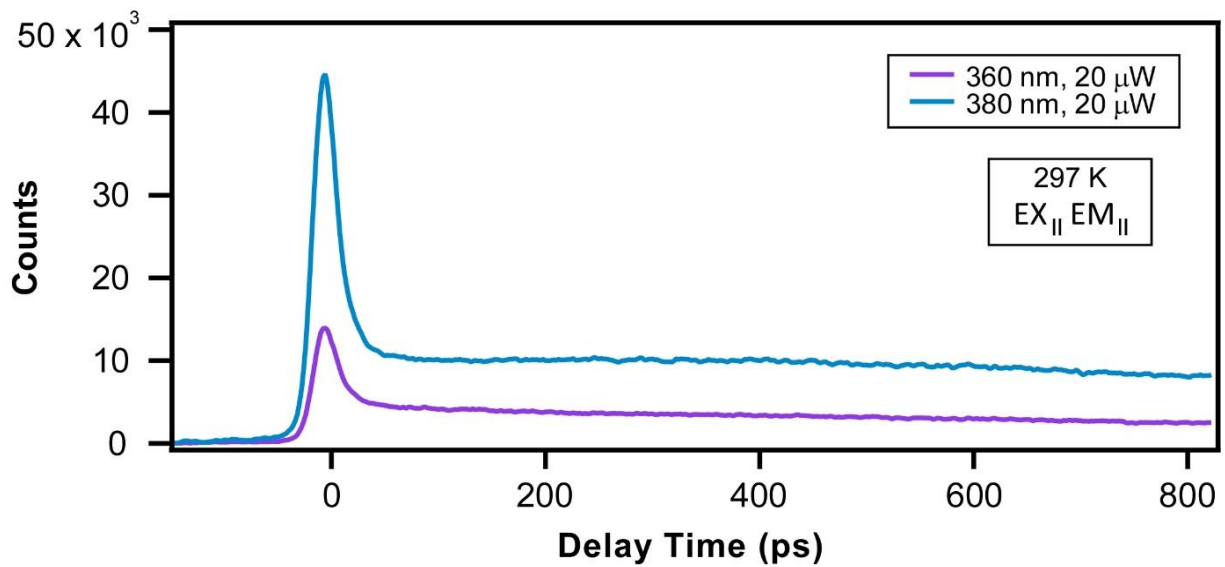

**Figure S3.** Comparison of PL counts as a function of delay time after 360 nm and 380 nm pulsed excitation at room temperature (Sample #5).

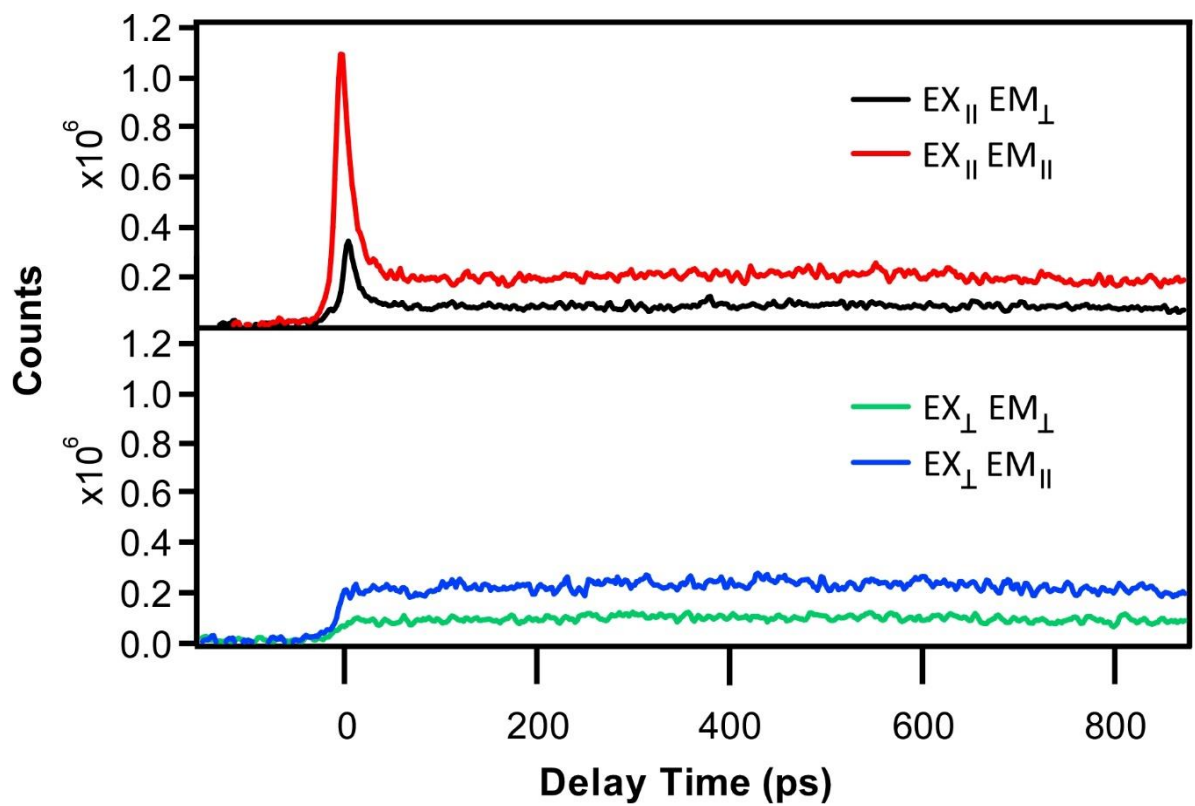

**Figure S4.** PL counts as a function of delay time after 380 nm pulsed excitation at room temperature (Sample #4) with different excitation and emission polarization configurations. The trend is similar to the data taken from Sample #5 shown in Figure 4 in the main text, though a small initial sharp decay peak is also observed for perpendicular emission polarization in this sample.

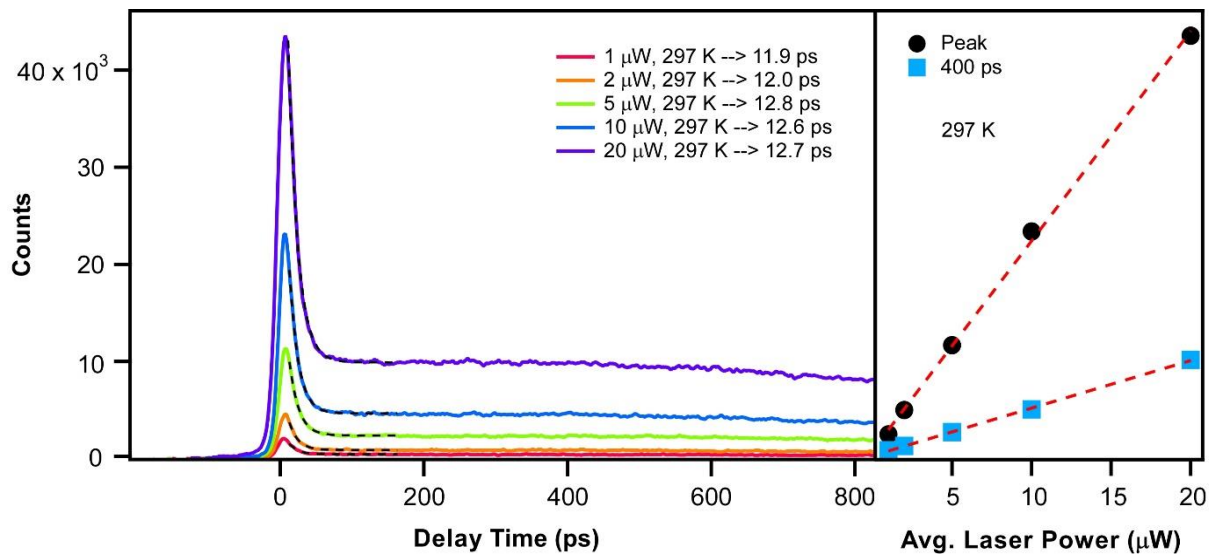

**Figure S5.** Left - PL counts as a function of delay time after 380 nm pulsed excitation at room temperature (Sample #5) under varying laser intensities. The fittings correspond to decay times for the sharp feature and are listed in the legend. Right - Total counts at peak as well as 400 ps later as a function of average laser power. Red dashed lines are linear fittings.

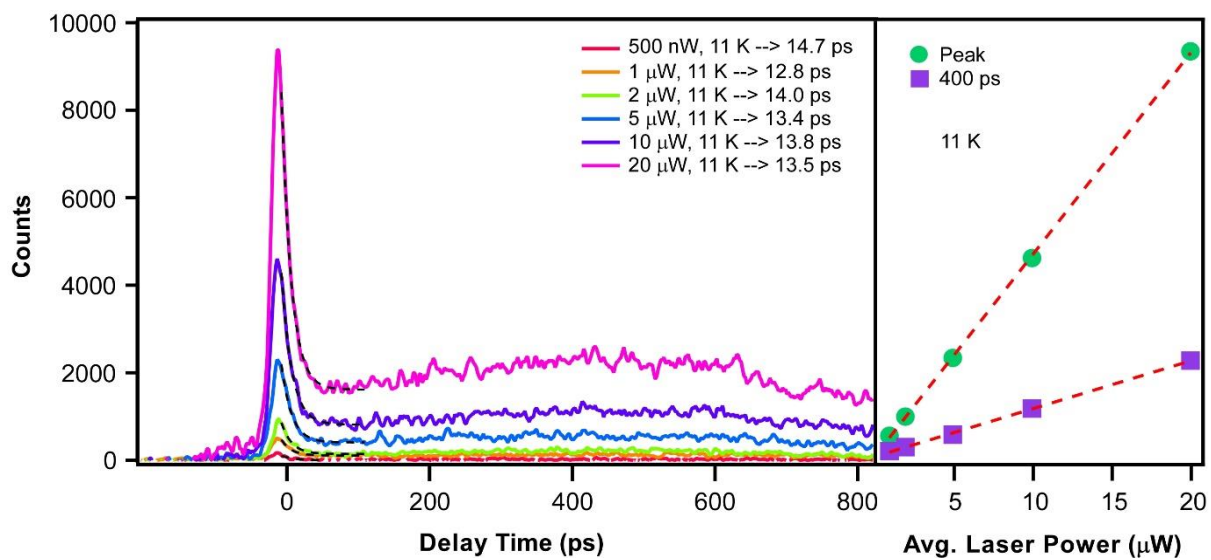

**Figure S6.** Left - PL counts as a function of delay time after 380 nm pulsed excitation at 11 K (Sample #5) under varying laser intensities. The fittings correspond to decay times for the sharp feature and are listed in the legend. Right - Total counts at peak as well as 400 ps later as a function of average laser power. Red dashed lines are linear fittings.